\documentclass[12pt]{iopart}

\newcommand{\rmf}{ {\rm f} }

\newcommand{\Mi}{M_\rmi}
\newcommand{\Mf}{M_\rmf}
\newcommand{\Li}{L_\rmi}
\newcommand{\Lf}{L_\rmf}

\newcommand{\bfA}{\bi{A}}
\newcommand{\bfr}{\bi{r}}
\newcommand{\bfp}{\bi{p}}
\newcommand{\bfL}{\bi{L}}
\newcommand{\bfx}{\bi{x}}
\newcommand{\bfv}{\bi{v}}
\newcommand{\bfN}{\bi{\nabla}}
\newcommand{\bfe}{\bi{e}}

\usepackage{graphicx}
\usepackage{url}

\begin{document}

\title{Transferring orbital and spin angular momenta of light to atoms}

\author{
A Pic\'{o}n$^{1}$\footnote{Present address: JILA, University of Colorado, 80309-0440 Boulder, USA.},
A Benseny$^{1}$,
J Mompart$^{1}$,
J R V{\'a}zquez de Aldana$^{2}$,
L Plaja$^{2}$,
G F Calvo$^{3}$ and
L Roso$^{4}$ }

\address{$^{1}$ Departament de F\'isica, Universitat Aut\`{o}noma de Barcelona, E-08193 Bellaterra, Spain}
\address{$^{2}$ Servicio L{\'a}ser, Universidad de Salamanca, E-37008 Salamanca, Spain}
\address{$^{3}$ Mathematics Department \& IMACI-Institute of Applied Mathematics in Science and Engineering, Universidad de Castilla-La Mancha, E-13071 Ciudad Real, Spain}
\address{$^{4}$ Centro de L\'aseres Pulsados, CLPU, Patio de Escuelas 1, E-37008 Salamanca, Spain}

\ead{antonio.picon@uab.cat}

\begin{abstract}
Light beams carrying orbital angular momentum, such as Laguerre--Gaussian beams, give rise to the violation of the standard dipolar selection rules during the interaction with matter yielding, in general, an exchange of angular momentum larger than $\hbar$ per absorbed photon. By means of \textit{ab initio} 3D numerical simulations, we investigate in detail the interaction of a hydrogen atom with intense Gaussian and Laguerre--Gaussian light pulses. We analyze the dependence of the angular momentum exchange with the polarization, the orbital angular momentum, and the carrier-envelope phase of light, as well as with the relative position between the atom and the light vortex. In addition, a quantum-trajectory approach based on the de Broglie--Bohm formulation of quantum mechanics is used to gain physical insight into the absorption of angular momentum by the hydrogen atom.
\end{abstract}



\maketitle

\section{Introduction}\label{sec:Intro}

In the past few years a great interest has been focused on helical light beams that are able to transport orbital angular momentum (OAM) and spin in its propagation direction~\cite{Allen, CalvoPRA} such as Laguerre--Gaussian (LG) beams. In fact, the transfer of OAM to matter has been already reported experimentally in Bose--Einstein condensates~\cite{Andersen} and ensambles of cold atoms~\cite{Kozuma}, and has been investigated for molecular systems~\cite{Nulty}. 
OAM has also attracted attention from the standpoint of applications ranging from optical tweezers~\cite{PadgettRev} to novel quantum information protocols~\cite{GrierNature}.  

At variance with previous works, where the OAM was coupled to the center of mass of an atomic ensamble, here we focus on the interaction of an OAM light pulse with a single atom allowing for electronic transitions to both bound states and continuum. In this context, we have recently derived novel selection rules for light-matter interactions~\cite{OptExPicon2010} where more than one $\hbar$ unit of angular momentum per photon are exchanged. Here, by means of 3D numerical simulations of the interaction between a hydrogen atom with an ultrashort and ultraintense light pulse carrying OAM, we investigate the angular momentum exchange in different photoionization scenarios and its manipulation through the light polarization (linear or circular), the carrier-envelope phase of the ultrashort pulse, and the relative position between the atom and the light vortex. 

Furthermore, we numerically calculate the de Broglie--Bohm quantum trajectories~\cite{Bohm-1,Bohm-2,Bohm-3} of the temporal evolution of the electron wavefunction to obtain physical insight into the electron dynamics. This formulation has already been applied in the strong field regime as a method to reduce the computational time to obtain the dynamics of a multielectron system~\cite{Christov-1,Christov-2} or to study the interaction of the hydrogen atom with an intense laser pulse in 1D or 2D, obtaining the high harmonic generation spectrum~\cite{Lai-HHG}, the above-threshold ionization energies of the electron~\cite{Lai-ATI}, as well as an insight on the role of the quantum potential in photoionization~\cite{Lai-Q}. However, to our knowledge, this is the first time that 3D photoionization is addressed with de Broglie--Bohm trajectories. Our motivation in using these trajectories focuses mainly on illustrating how the electron absorbs the angular momentum due to the light polarization (photon spin) and due to its transverse profile (OAM).

The article is organized as follows. In \sref{sec:Model} we introduce the physical system consisting of a hydrogen atom interacting with an ultrashort light pulse described in the LG basis.
We proceed investigating different photoionization scenarios, starting with the standard linearly and circularly polarized Gaussian pulses, \sref{sec:Gaussian}.
We then compare these results with the ones with LG pulses, \sref{sec:LaguerreGaussian}, with linear or circular polarization, where the combination of spin and OAM is studied.
We then investigate the role of the carrier-envelope phase and the case where the LG vortex is not centered with respect to the atom position in sections \ref{sec:CEP} and \ref{sec:NotCentered}, respectively.
In \sref{sec:LongPulses}, we examine longer light pulses, where trapping of the electron state by the light pulse is observed.
Finally, we present the conclusions in \sref{sec:Conclusions}.

\section{Physical model}\label{sec:Model}

\begin{figure}
\begin{center}
\includegraphics[width = 0.6 \textwidth]{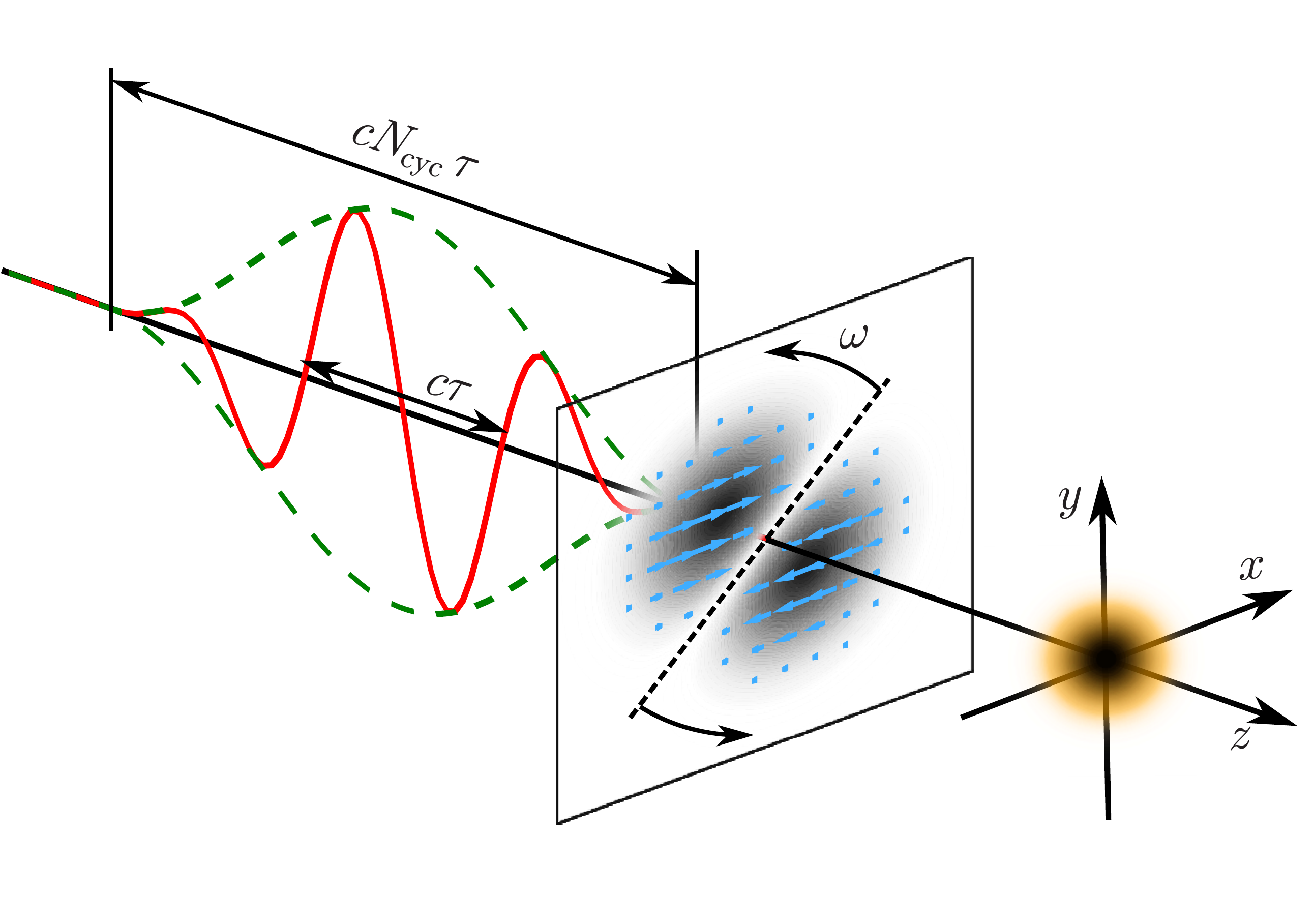}
\end{center}
\caption{\label{fig:Sketch}
(colour online) Sketch of the physical system consisting of a light pulse interacting with a hydrogen atom at the origin, where the time pulse dependence and its associated initial parameters are introduced.
As an example, we plot the transverse profile of the vector potential \eref{Eq:VectorPotential} for a linearly polarized LG mode with topological charge $\ell = 1$.}
\end{figure}

The system under investigation is sketched in \fref{fig:Sketch}: a hydrogen atom interacts with an ultrashort and ultraintense light pulse carrying OAM.
Being the nucleus a massive object, we can assume that the dynamics are given mainly by the electron excitation.
The non-relativistic Schr{\"o}dinger equation describing the dynamics of the electron in interaction with the light field reads: 
\begin{eqnarray}\label{Eq:Schrodinger}
\rmi \hbar \frac{\partial}{\partial t} \psi(\bfr, t)  = \mathcal{H} \psi(\bfr, t)  = \left[ \frac{1}{2m} \left( \bfp - q \,\bfA (\bfr, t) \right)^{2} + q \, V(r)\right] \psi(\bfr, t) \; ,
\end{eqnarray}
being $\psi(\bfr, t)$ the electron quantum state at time $t$, $V(r)$ the Coulomb potential due to the interaction with the nucleus, $m$ the electron mass, $q$ the electron charge, $\bfp$ the linear momentum operator, and $\bfA(\bfr, t)$ the vector potential of the light electromagnetic field. 

\subsection{Ultrashort light pulse}\label{sec:LightPulse}

We consider an ultra-short laser pulse carrying OAM, whose spatial profile will be expressed as an LG mode~\cite{Allen,CalvoPRA}. Considering the wave propagating along the $z$ direction, these LG modes read
\begin{eqnarray}
\eqalign{
LG_{\ell,p}(\rho,\phi,z)=&\sqrt{\frac{2p!}{\pi (\vert \ell\vert + p)!}}\frac{w_{0}}{w(z)} L_{p}^{\vert \ell\vert}\!\left(\frac{2\rho^{2}}{w^{2}(z)}\right) \left(\frac{\sqrt{2}\rho}{w(z)}\right)^{\!\vert \ell\vert}\! \\ &\times \exp\left(-\frac{\rho^{2}}{w^{2}(z)}\right)\exp \left(i\ell\phi+i\frac{k_{0}\rho^{2}}{2R(z)}+i\Phi_{\rm G}(z)\right)\, , }
\end{eqnarray}
where $\rho=\sqrt{x^{2}+y^{2}}$, $\phi = \arctan(y/x)$, $w(z)=w_{0}\sqrt{1+(z/z_{0})^{2}}$ is the beam width, $w_{0}$ is the beam waist at $z =0$, $z_{0}=k_{0}w^{2}_{0}/2$ is the Rayleigh range, $k_{0}$ is the carrier wave number, $R(z)=z\left(1+(z_{0}/z)^{2}\right)$ is the phase-front radius, $\Phi_{\rm G}(z)=-(2p+\vert \ell\vert +1)\arctan(z/z_{0})$ is the Gouy phase, and $L_{p}^{\vert \ell\vert}(\xi)$ are the generalized Laguerre polynomials
\begin{eqnarray}
L_{p}^{\vert \ell\vert}(\xi) = \sum_{m=0}^{p} (-1)^{m}\frac{(\vert \ell\vert + p)!}{(p-m)!\,(\vert \ell\vert + m)!\,m!}\,\xi^{m}.
\label{Eq:Laguerre}
\end{eqnarray}
The two independent indices $\ell=0,\pm 1,\pm 2,\ldots$ and $p=0, 1, 2,\ldots$ correspond to the topological charge and the number of nonaxial radial nodes of the mode.  Note that for $\ell = p = 0$ we recover the standard Gaussian beam. As shown in Refs.~\cite{Allen,CalvoPRA}, in the paraxial regime LG modes carry a discrete OAM of $\ell\hbar$ units per photon along their propagation direction. 

The temporal envelope of our light pulse, see \fref{fig:Sketch}, is parametrized by a sine squared function, with frequency $\omega_{\rme} = \pi/N_{\rm cyc}\tau$, where $N_{\rm cyc}$ and $\tau$ are the cycle number and the period of the carrier wave, respectively. Hence, the vector potential can be cast in the form 
\begin{eqnarray}\label{Eq:VectorPotential}
\eqalign{
\bfA(\bfr,t)=& \bfA_{0} \, \sin^{2}\left(\frac{\omega_{\rme}}{c}(z+a_{0} )-\omega_{\rme}t \right) \rme^{\rmi\frac{\omega}{c} (z-ct)+i\chi}\;LG_{\ell,p}(\rho,\phi,z) \\ & \times \left(\theta_H(z-ct+\pi c/\omega_{\rme} + a_{0})-\theta_H(z-ct+a_{0}) \right) + {\rm  c.c.}, }
\end{eqnarray}
being $\theta_H(x)$ the Heaviside or step function, $a_{0}$ the Bohr radius, $c$ the speed of light, $\chi$ the `global' carrier-envelope phase,  $\omega = c k_0 = 2\pi / \tau$ the frequency of the carrier wave, and $\bfA_{0}$ the vector amplitude of the wave.
In the Coulomb gauge, the vector potential has a small polarization component in the propagation direction~\cite{Allen}, which has been neglected here.
Note that the wavefront is helicoidal.

The polarization in the $xy$ plane is included in $\bfA_{0} \equiv A_{0} (\alpha \, \bfe_x + \beta \, \bfe_y)$, where $\alpha$ and $\beta$ are complex numbers satisfying $\vert \alpha \vert^{2} + \vert \beta \vert^{2} = 1$.
For example, for a circular polarization, $\alpha=1$ and $\beta = i s$. Note that here we make use of the notation of~\cite{PolarizationNotation}, where a left (right) circularly polarized beam carries angular momentum $s \hbar$ with $s=+1$ ($s=-1$) along the propagation direction.
A linear polarization forming an angle $\phi$ with the $x$ axis corresponds to $\alpha = \cos \phi$ and $\beta = \sin \phi$ that is an equal superposition of the two circular polarizations.
We would like to emphasize the fact that the electric field amplitude and the LG `local' carrier-envelope phase that interacts with each part of the electron wavefunction depend on the transverse position, at variance with the plane wave case.
Note that since our description of the electromagnetic field is in terms of the vector potential $\bfA$, it includes the effects of both electric and magnetic fields of the incident laser pulse.

\subsection{Selection rules}\label{sec:SelectionRules}

Hamiltonian \eref{Eq:Schrodinger} of the system can be split as $\mathcal{H}=\mathcal{H}_0+\mathcal{H}_I+\mathcal{H}_{II}$, where $\mathcal{H}_0$ is the free hydrogen Hamiltonian (kinetic and Coulomb terms), $\mathcal{H}_I\equiv -q \, (\bfp\cdot\bfA+\bfA\cdot\bfp)/2m$ and $\mathcal{H}_{II}\equiv q^{2}\bfA^{2} /2m$.  The selection rules associated to $\mathcal{H}_I$ are (see Appendix A):
\begin{eqnarray} \label{Eq:SelectionRulesH1}
\vert\Delta L\vert \leq \vert\ell\vert + 1 \leq \Li + \Lf , \quad \Delta M = \pm (\ell + s), \quad \Delta L + \ell \hspace{0,2cm} \textrm{is odd} .
\end{eqnarray}
It comes clear from \eref{Eq:SelectionRulesH1} that for $\ell=0$ we recover the usual selection rules for plane waves.  Furthermore, for $\ell \neq 0$ we can excite transitions with an angular momentum exchange of more than one unit of $\hbar$. Selection rules~\eref{Eq:SelectionRulesH1}, in photon terms, can be thought as the absorption of a photon carrying a total angular momentum $\ell + s$ in the propagation direction, where $s$ accounts for the spin angular momentum.

In the case of plane waves, $\mathcal{H}_{II}$ is uniform in the transverse plane, yielding just a ponderomotive force but, interestingly, for $\ell \neq 0$ it can produce transitions between spherical harmonics.
In fact, proceeding in a similar manner as with $\mathcal{H}_I$, one obtains~\cite{OptExPicon2010}:
\begin{eqnarray} \label{Eq:SelectionRulesH2}
\vert\Delta L\vert \leq 2 \vert\ell\vert \leq \Li + \Lf , \quad \Delta M = \pm 2 \ell , \quad \Delta L \hspace{0,2cm} \textrm{is even} .
\end{eqnarray}
Since selection rules~\eref{Eq:SelectionRulesH2} appear from a term with $\bfA^{2}$ we expect them to be associated with two-photon processes, and, therefore will be less significant than those associated with selection rules~\eref{Eq:SelectionRulesH1}, since the intensity near the vortex, where the atom is located, is relatively small.

In the following section, where the complete dynamics of the system are numerically solved, these selection rules will be essential to understand the interaction process. We would like to remark that these selection rules apply to scenarios beyond photoionization, such as transitions between bound states.

\section{Numerical simulations}\label{sec:NumericalSimulations}

Initially, before the interaction with the light pulse, we assume the electron to be in the hydrogen's ground state, i.e., $\psi_{0} (\bfr)= \psi (\bfr,t=0)= 1/\sqrt{\pi a_{0}^{3}}\rme^{-r/a_{0}}$.
We will first consider the atom to be centered in the light propagation axis, i.e., at $\rho = 0$, corresponding to the maximum amplitude of the Gaussian profile or the LG vortex position, and a pulse carrier-envelope phase $\chi = 0$.
Eventually, we will assume arrangements where the atom is not centered in the light vortex and study the role of the carrier-envelope phase.

In all the following cases we consider the pulse, see~\eref{Eq:VectorPotential}, to have $N_{\rm cyc}=3$ (except in \sref{sec:LongPulses}), a carrier frequency of $\omega = 1$~au ($2\pi \times 6.57\cdot10^{15}$ s$^{-1}$, ultraviolet) and period $\tau = 2\pi$~au ($152$ as)\footnote{Atomic units (au), where $\hbar=m=q=1/(4\pi\epsilon_0)=1$, are used throughout the paper. The length unit is $a_0$ and the unit of angular momentum is $\hbar$.}. We assume a beam waist that satisfies the paraxial regime, $w_{0}=9\cdot10^{4}$~au ($4.79$ $\mu$m), which is much larger than the typical size of the atom ($w_{0} \gg a_{0}$). Note that in this regime, the Gaussian case is equivalent to a plane wave. For simplicity, since the OAM of the light is given only by the topological charge $\ell$, we will restrict ourselves to LG modes with $p=0$.

We express the electron quantum state at each time as $\vert \psi (t)\rangle = c_0\, \vert \psi_{0}\rangle + \vert \delta\psi (t)\rangle$, where $\vert \delta\psi \rangle$ is the excited part of the state and $c_0 = \langle \psi_{0} \vert \psi (t)\rangle$.
Since the field has a frequency $\omega=1$~au, larger than the bound hydrogen energy (0.5~au), we assume that, at the end of the pulse, the excited part will mainly account for ionization. Thus, we define the ionization probability as $P_{\rm I} = \vert \langle \delta \psi (t) \vert \delta \psi (t)\rangle \vert ^2 = 1-\vert c_0 \vert^2$.
To clearly observe the effects of the light on the atom, we will only plot in the figures the probability density of the excited part of the wavefunction.

By means of the Crank--Nicolson algorithm, we integrate Schr\"odinger equation~\eref{Eq:Schrodinger} in a 3D grid of $N_x \times N_y \times N_z = 400 \times 400 \times 250$ points, covering a space of $(-20, \, 20)$~au $\times$ $(-20, \, 20)$~au in the polarization plane and of $(-12.5, \, 12.5)$~au in the propagation direction.
We take absorbing boundary conditions in order to avoid unphysical reflections when the ionized wavefunction reaches the borders of the integration grid.
The decrease in the norm of the wavefunction due to these absorbing boundaries will be in most cases relatively small (less than 2 \%) and as usually, it will be regarded as ionized population with kinetic energy large enough to reach the boundaries.

During the evolution we will calculate average values for different observables such as the position of the electron $\langle \bfr \rangle$, or its angular momentum $\langle \bfL \rangle =  \langle \bfr \times (\bfp - q\,\bfA) \rangle$.
Moreover, we will project numerically the quantum state onto the spherical harmonics in order to check the agreement with the previously presented selection rules.
To acquire physical insight on the dynamics, we will compute the de Broglie--Bohm quantum trajectories from the evolution of the quantum state (see Appendix B for details of the quantum trajectories formulation). To ease the reading, from now on we will refer to them simply as trajectories.

\subsection{Gaussian pulse}\label{sec:Gaussian}

In this section, we consider a Gaussian pulse, i.e., with $\ell = 0$, $p=0$. We take the electric field amplitude to be $A_{0} \, \omega = 0.33$~au (corresponding to a peak intensity of $3.9\cdot10^{15}$~W/cm$^2$) for both cases of linear and circular polarization.

\subsubsection{Linear polarization}\label{sec:GaussianLinear}

\begin{figure}
\begin{center}
\includegraphics[width = 0.95 \textwidth]{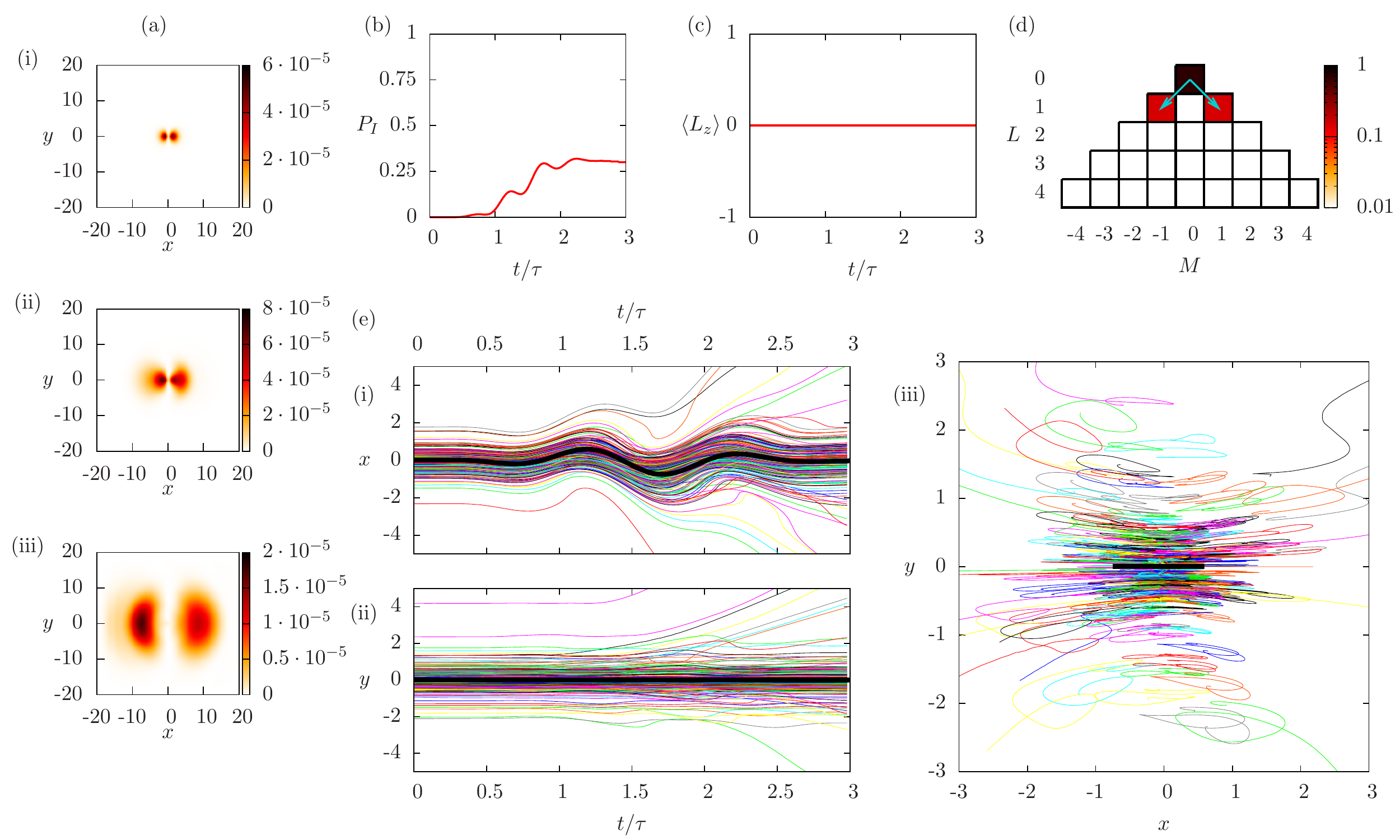}
\end{center}
\caption{\label{fig:GaussLin}
(colour online) Results of a simulation with a Gaussian pulse ($\ell = 0$, $p=0$) linearly polarized in the $x$ direction with electric field amplitude $A_{0} \, \omega = 0.33$~au. For the rest of the parameters, see text.
(a) Projection of the excited state $\vert \delta \psi (t)\rangle$ onto the plane $xy$ at times $t=\tau$, $t=2 \tau$, and $t=3 \tau$.
Temporal evolution of 
(b) the ionization probability $P_{\rm I}$ and
(c) the expected value of the angular momentum along the $z$ axis, $\langle L_z \rangle$.
(d) Projection of the electron quantum state at the end of the pulse into spherical harmonics $Y^{M}_{L}$. Arrows correspond to transitions allowed by selection rules shown in \sref{sec:SelectionRules}.
(e) Projection of the 3D de Broglie--Bohm trajectories onto the $x$ and $y$ axis over time and in the $xy$ plane. The thick black curves in (e) correspond to the mean value of the electron position.}
\end{figure}

The results of the simulation with a pulse polarized in the $x$ direction, plotted in \fref{fig:GaussLin}, show that the electron begins to be ionized in the first cycle and, at the end of the pulse, 30\% of the quantum electron state is ionized, see \fref{fig:GaussLin}(b).
We also see in \fref{fig:GaussLin}(c) that there is no exchange of angular momentum in the $z$ direction since both the light pulse is linearly polarized and does not carry OAM.
\Fref{fig:GaussLin}(d) shows that the most excited spherical harmonics are $Y^{1}_{1}$ and $Y^{-1}_{1}$, in complete agreement with the selection rules discussed in \sref{sec:SelectionRules} since the linear polarization is a superposition of left and right circular polarizations. 

The motion of the electron wavepacket is unravelled in the trajectories plotted in \fref{fig:GaussLin}(e). Trajectories describe oscillations at the field frequency along the polarization axis following (with a $\pi/2$ phase delay) the oscillation of the electric field.
Some of the trajectories, i.e., the ones that start far enough from the nucleus, are ionized near the electric field maxima.
The mean value of the position oscillates in the $x$ axis while remains in the origin of the $yz$ plane, except for a small shift ($\sim 10^{-2}$~au) on the $z$ direction due to the magnetic field, in agreement with previous results~\cite{Javi}. A shift around this value will appear in all the following scenarios.

The Coulomb potential acting over the trajectories that are closer to the nucleus retains them in the nucleus vicinity. The ones that distance from it make the trajectory cloud expand. This can also be seen as the spreading of the quantum state function away from the nucleus where the Coulomb potential vanishes, see \fref{fig:GaussLin}(a).

\subsubsection{Circular polarization}\label{sec:GaussianCircular}

\begin{figure}
\begin{center}
\includegraphics[width = 0.95 \textwidth]{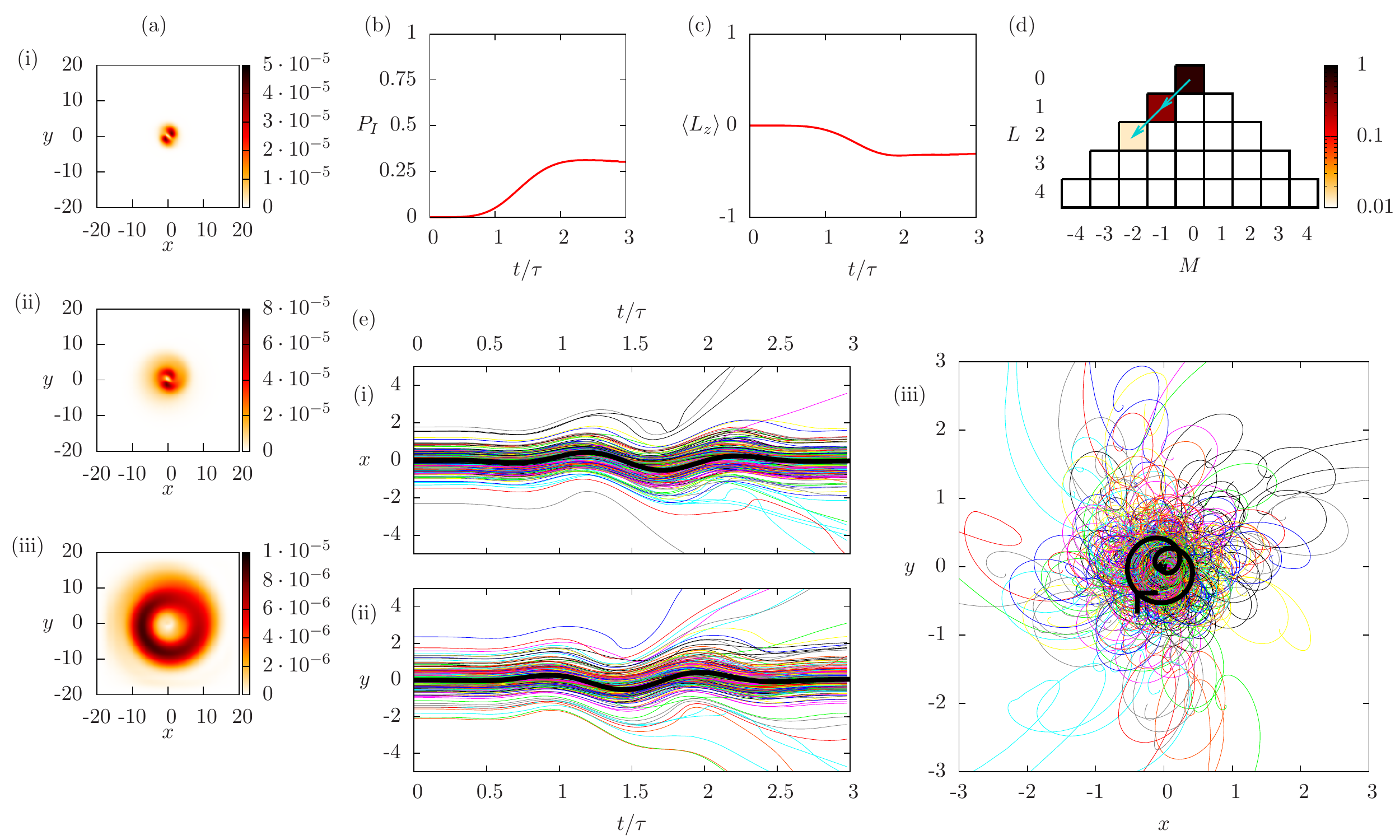}
\end{center}
\caption{\label{fig:GaussRCirc}
(colour online) The same as for \fref{fig:GaussLin} but with right circular polarization ($s=-1$). The arrow in (e-iii) indicates the rotation direction of the position mean value.}
\end{figure}

In \fref{fig:GaussRCirc} we present the results of a simulation with a right circularly polarized pulse. We have checked that in the case of left circular polarization the results of the simulation are completely symmetric to the ones discussed below.

As in the case of linear polarization, the ionization probability reaches about 30\% at the end of the pulse, see \fref{fig:GaussRCirc}.
At variance with the linearly polarized Gaussian pulse, which does not transfer angular momentum in the $z$ direction to the electron, in \fref{fig:GaussRCirc}(c) we see that, after the interaction, the electron has $\langle L_{z} \rangle =-0.3$~au.
As selection rules predict, we expect that a right (left) circularly polarized light mostly excites the $Y_{1}^{-1}$ ($Y_{1}^{1}$) spherical harmonic, which is in accordance with the numerical results in \fref{fig:GaussRCirc}(d).
Note that the electron is excited to $L = 1$ states through one-photon processes and to $L = 2$ states through two-photon processes.

The trajectories depicted in \fref{fig:GaussRCirc}(e) describe oscillations along the two polarization axis.
It is noteworthy that for each of the trajectories, the main dynamics in these two axis are independent.
The mean value of the electron position shows that the electric field displaces the entire trajectory cloud from the propagation axis and makes it rotate clockwise (since it is right circularly polarized) around its initial position.
This is due to the fact that the electric field is almost uniform in the transverse plane, therefore acting in the same way over all the trajectories.

\subsection{Laguerre--Gaussian pulse}\label{sec:LaguerreGaussian}

From now on we consider pulses bearing OAM, e.g., with $\ell = 1$ and $p=0$, for both the linear and circular polarization cases.

We consider our atom to be centered at the light vortex, where the electric field amplitude is zero and increases linearly in the vicinity of the singularity.
Thus, we require very intense lasers to affect the atom. We choose a pulse with $A_{0} \, \omega = 1.4 \cdot 10^{4}$~au, corresponding to a peak intensity of $6.7\cdot10^{24}$~W/cm$^2$, although the electric field at a distance of 1~au from the vortex is only 0.16~au.
Albeit the considered light intensity is unreachable nowadays, the reader must take into account that very strong lasers are under construction \cite{ELI-web}, planning to reach up to $10^{26}$~W/cm$^2$ at 800~nm wavelength. 

\subsubsection{Linear polarization}\label{sec:LGLinear}

\begin{figure}
\begin{center}
\includegraphics[width = 0.95 \textwidth]{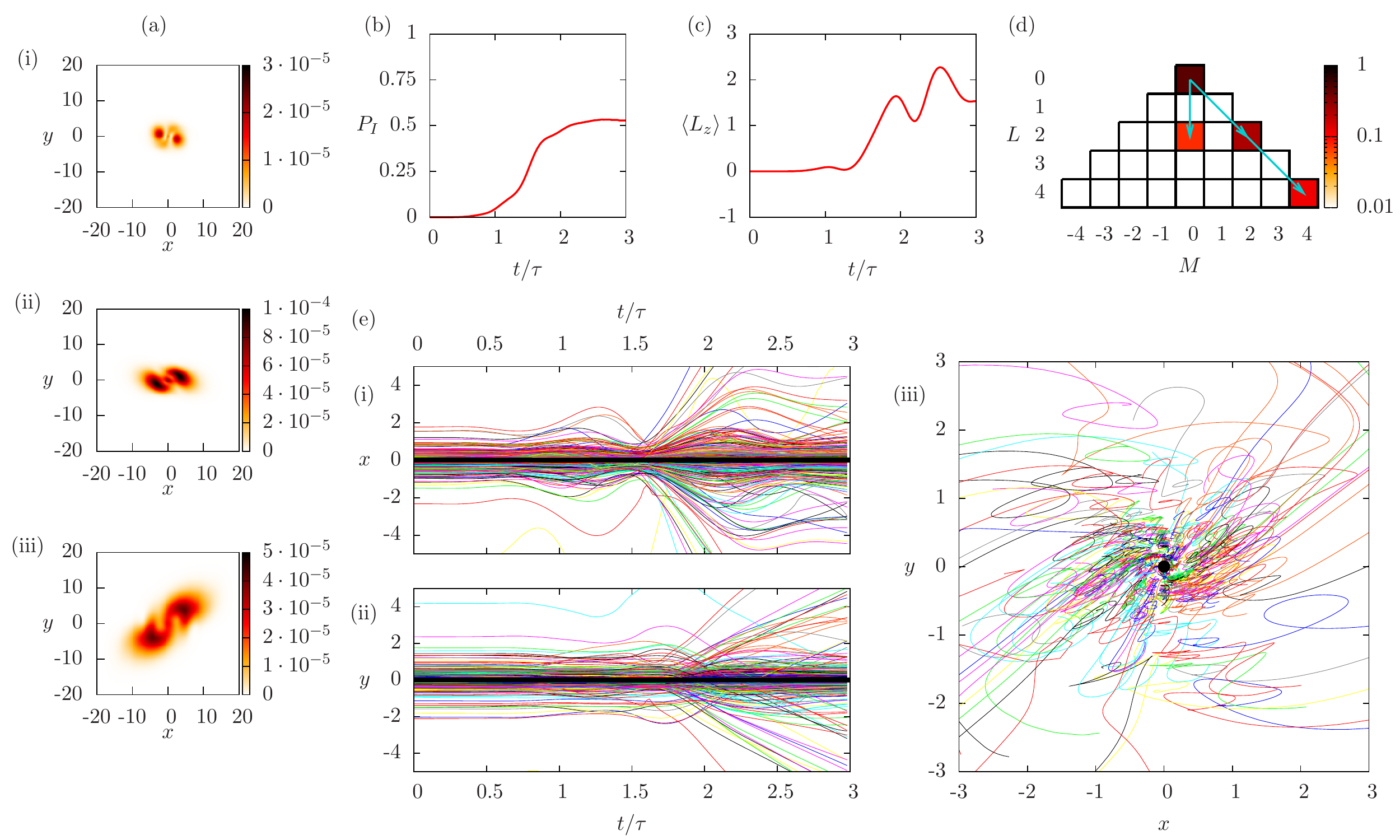}
\end{center}
\caption{\label{fig:LGLinear}
(colour online) The same as for \fref{fig:GaussLin} but with an LG mode with $\ell=1$ linearly polarized in the $x$ direction with electric field amplitude $A_{0} \, \omega = 1.4 \cdot 10^{4}$~au. }
\end{figure}

In \fref{fig:LGLinear} we show the results of a simulation with an LG pulse with $\ell=1$ polarized in the $x$ direction.
We have checked that due to the symmetry of the problem, the interaction with an $\ell=-1$ pulse leads to the expected symmetric results.

As shown in \fref{fig:LGLinear}(a), during the interaction with the pulse the electron is ionized, but it remains trapped around the light vortex, due to the ponderomotive potential of the pulse.
Once the electric field intensity decreases, the electron state begins to expand, as it can be seen in \fref{fig:LGLinear}(a-iii) or the trajectories depicted in \fref{fig:LGLinear}(e).
This trapping effect will be studied in more detail in \sref{sec:LongPulses}.
Note that even though the pulse is linearly polarized in the $x$ direction, the OAM breaks the symmetry on the $y$ axis.
At the end of the pulse the ionization probability is 53 \%.

It is interesting to point out the total angular momentum transferred to the electron, see \fref{fig:LGLinear}(c). The electron starts in the ground state, with zero angular momentum and, as the pulse interacts with the electron, it oscillates, reaching, at the end of the pulse, a finite amount of angular momentum of 1.53~au, at variance with the linearly polarized Gaussian case where no exchange of angular momentum took place.
Thus, we expect that the populated electron excited states at the end of the pulse bear angular momentum.
The spherical harmonics populations plotted in \fref{fig:LGLinear}(d) show perfect agreement with the selection rules from \sref{sec:SelectionRules}, since the linear polarization is a superposition of the two circular polarizations (see next section for further details), i.e., $\Delta M = \ell \pm 1$.
Note that the electron is excited to $L = 2$ states through one-photon processes and to $L = 4$ states through two-photon processes.

The mean motion of the electron state is negligible. Nevertheless, the dynamics are far from simple, due to the electric field having a spatial profile where it points in opposite directions in opposite points of the polarization plane. 
A peculiar effect is that, even though the polarization is linear in the $x$ direction, the trajectories are forced to acquire a velocity in the $y$ direction, in order to rotate counterclockwise, due to the light $\ell = 1$ OAM, giving rise to the absorption of angular momentum by the electron.

Notice that in the circularly polarized Gaussian case, the rotation of the trajectories around the propagation axis took place in sequence with each other, whereas here each of the trajectories rotates independently around the axis.

\subsubsection{Circular polarization}\label{sec:LGCircular}

\begin{figure}
\begin{center}
\includegraphics[width = 0.95 \textwidth]{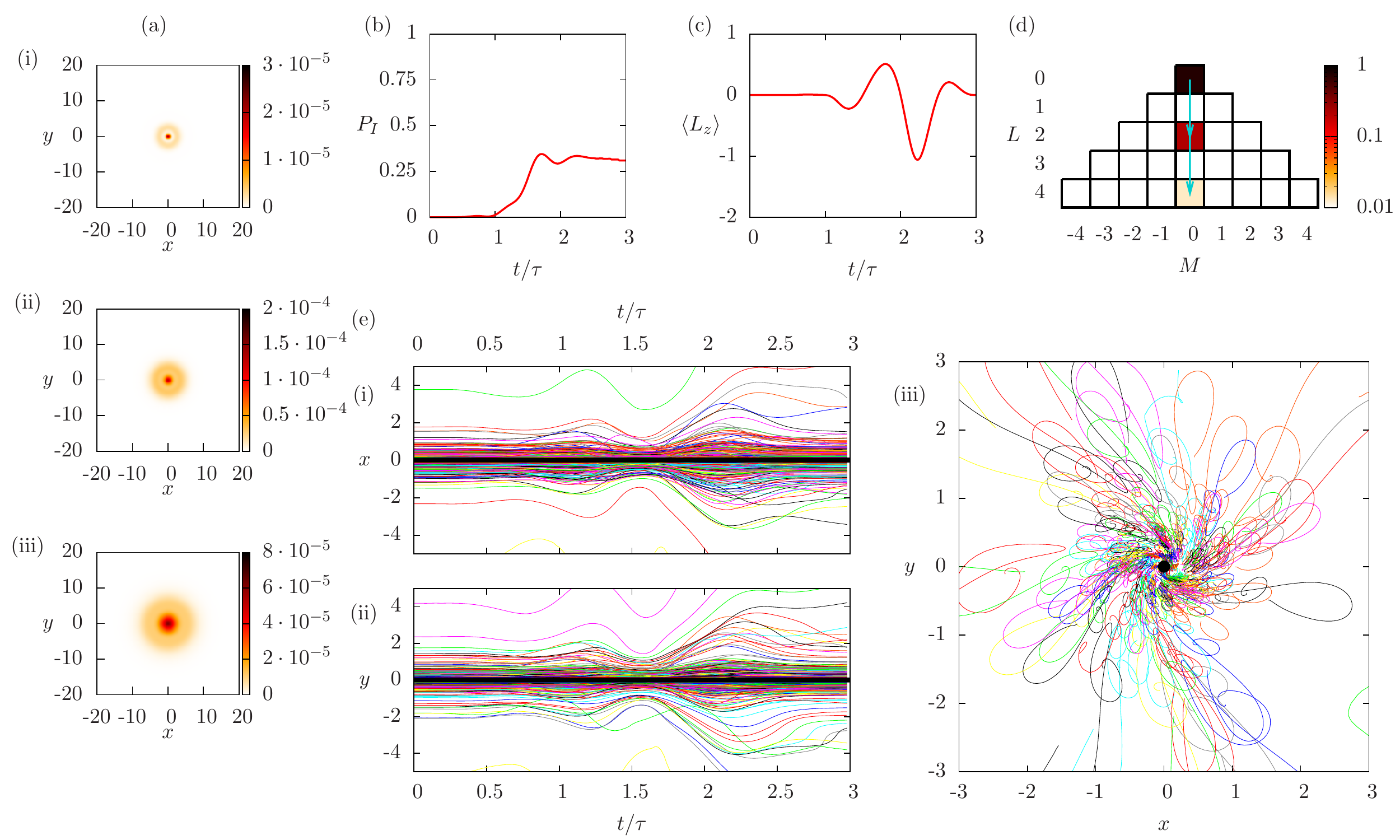}
\end{center}
\caption{\label{fig:LGRCircular}
(colour online) The same as for \fref{fig:LGLinear} but with right circular polarization ($s=-1$).}
\end{figure}
\begin{figure}
\begin{center}
\includegraphics[width = 0.95 \textwidth]{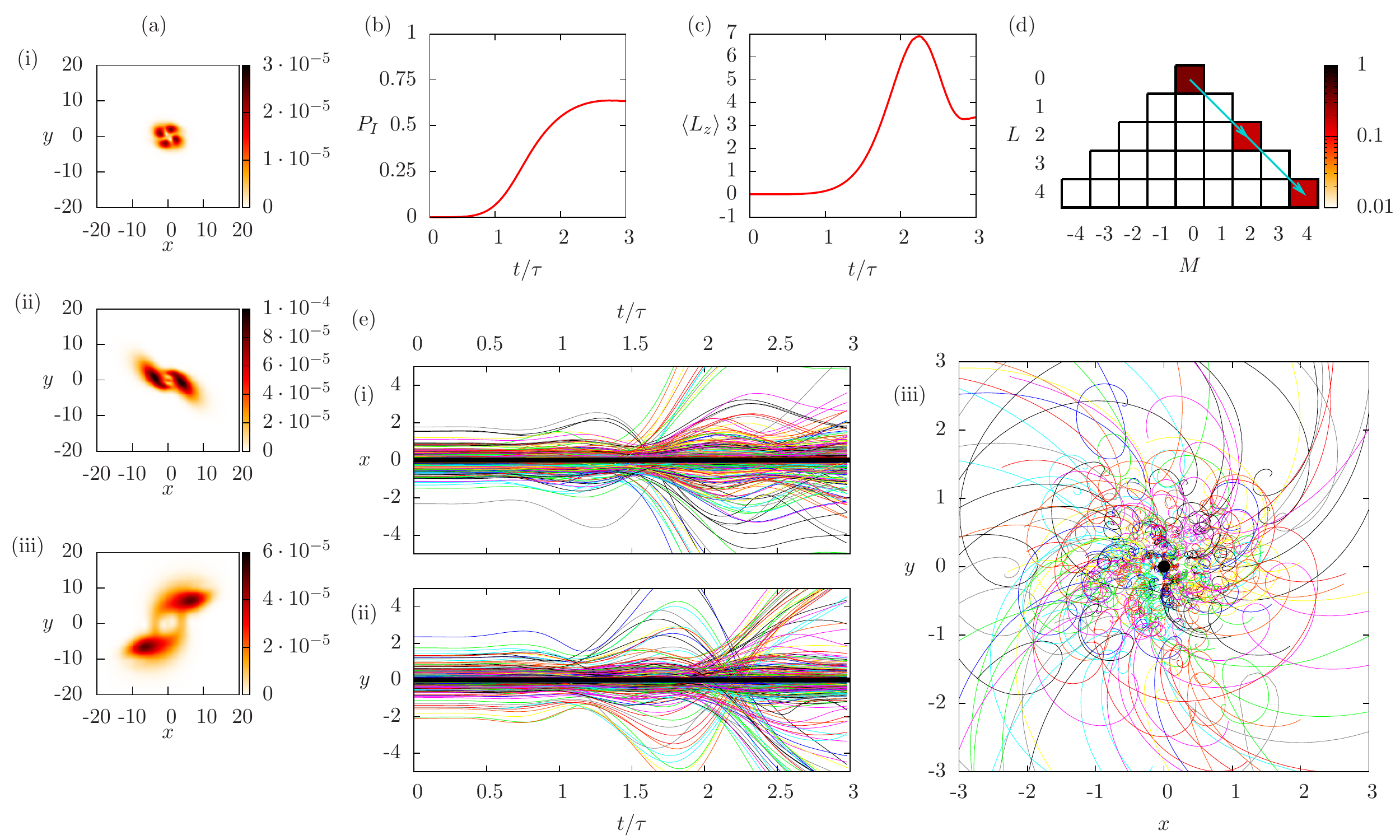}
\end{center}
\caption{\label{fig:LGLCircular}
(colour online) The same as for \fref{fig:LGLinear} but with left circular polarization ($s=+1$).}
\end{figure}

In this section we consider the same pulse as in \sref{sec:LGLinear} ($\ell = 1$), but circularly polarized.
At variance with \sref{sec:GaussianCircular}, the dynamics for the left and right circular polarization cases are different due to the relative sign between the light OAM and polarization components of the angular momentum along the $z$ direction.
This comes clear from the results shown in figures~\ref{fig:LGRCircular}(a) and \ref{fig:LGLCircular}(a), whereas the right circular polarization preserves the circular symmetry of the excited quantum state, the left circular polarization gives rise to a more complex behaviour due to the exchange of angular momentum.
Note that in the case of right circular polarization ($s = -1$) the spin and OAM angular components compensate while in the left circular polarization case ($s = 1$) they add up.

Figures~\ref{fig:LGRCircular}(b) and \ref{fig:LGLCircular}(b) show that, even though the electric field intensity is the same for both cases, in the left circular polarization case the ionization is approximately twice that in the case of right circular polarization (63 \% and 31 \%, respectively). In this particular case, we attribute this enhancement of the ionization to the exchange of OAM, as can be seen comparing figures~\ref{fig:LGRCircular}(c) and \ref{fig:LGLCircular}(c). At the end of the pulse the expected value for $L_z$ in the left circular polarization case is 3.35~au, while becomes null for the right circular polarization.

The transfer of angular momentum to the electron is clearly seen in figures~\ref{fig:LGRCircular}(d) and \ref{fig:LGLCircular}(d), in perfect agreement with the previously discussed selection rules.
We can check that a right circular polarization excites states with $M=0$, as $\Delta M = \ell + s = 0$, therefore not transfering any OAM in the $z$ direction to the atom.
On the other hand, a left circular polarization excites states obeying $\Delta M = \Delta L = \ell + s = 2$. 

As in the linear polarization case, the main dynamics are concentrated on the polarization plane and the mean motion of the electron state is again negligible. The absorption of angular momentum by the electron is clearly depicted in the trajectories: while in the left circular polarization the electric field of the light exerts a torque on the trajectory cloud forcing it to rotate around the vortex, see \fref{fig:LGLCircular}(e-iii), the combination of OAM and right circular polarization partially inhibits this rotation see \fref{fig:LGRCircular}(e-iii).

\subsection{Carrier-envelope phase manipulation}\label{sec:CEP}

\begin{figure}
\begin{center}
\includegraphics{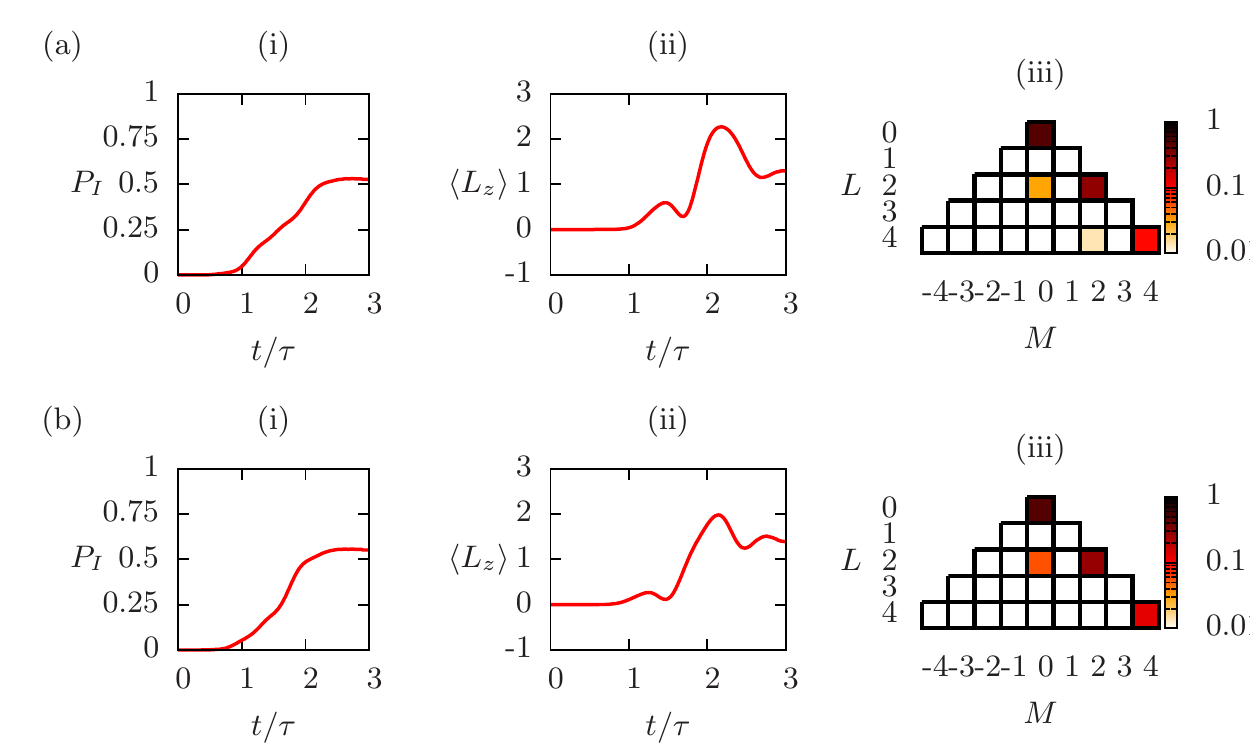}
\end{center}
\caption{\label{fig:CEP}
(colour online) Results of a simulation with an LG pulse ($\ell=1$, $p=0$) linearly polarized in the $x$ direction with electric field amplitude $A_{0} \, \omega = 1.4 \cdot 10^{4}$~au, for a carrier-envelope phase of (a) $\chi = \pi$ and (b) $\chi = \frac{\pi}{2}$.
Temporal evolution of (i) the ionization probability $P_{\rm I}$ and (ii) the expected value of the angular momentum along the $z$ axis, $\langle L_z \rangle$. (iii) Projection of the electron quantum state at the end of the pulse into spherical harmonics $Y_L^M$. }
\end{figure}

Here we consider the same pulse that in \sref{sec:LGLinear} (a linearly polarized $\ell=1$ LG pulse), but modifying the carrier-envelope phase (CEP) $\chi$, see expression of the vector potential \eref{Eq:VectorPotential}.
In \fref{fig:CEP} we show the results for two different values of the CEP (a) $\chi=\pi$ and (b) $\chi=\pi/2$.
Now, the ionization reaches 52\% for $\chi = \pi$, and 55\% for $\chi = \pi/2$, very similar to the pulse of \sref{sec:LGLinear}, of 53 \%.
The fact that the ionization probability does not significantly depend on the CEP for linearly polarized LG pulses, can be understood by noticing that the `local' CEP that interacts with each part of the electron wavefunction takes all possible values due to its azimuthal phase dependence.

The final angular momentum is 1.30~au when $\chi = \pi$, and 1.39~au when $\chi = \pi/2$, relatively smaller than the 1.53~au obtained obtained for $\chi = 0$.
The fact that the absorbed angular momentum by the electron slightly varies with the CEP affects the probability of each excited spherical harmonic, but not the states permitted by the selection rules, which are the same in all cases, as can be seen comparing figures~\ref{fig:CEP}(a-iii), \ref{fig:CEP}(b-iii) and \ref{fig:LGLinear}(d).

\subsection{Off-center vortex}\label{sec:NotCentered}

\begin{figure}
\begin{center}
\includegraphics{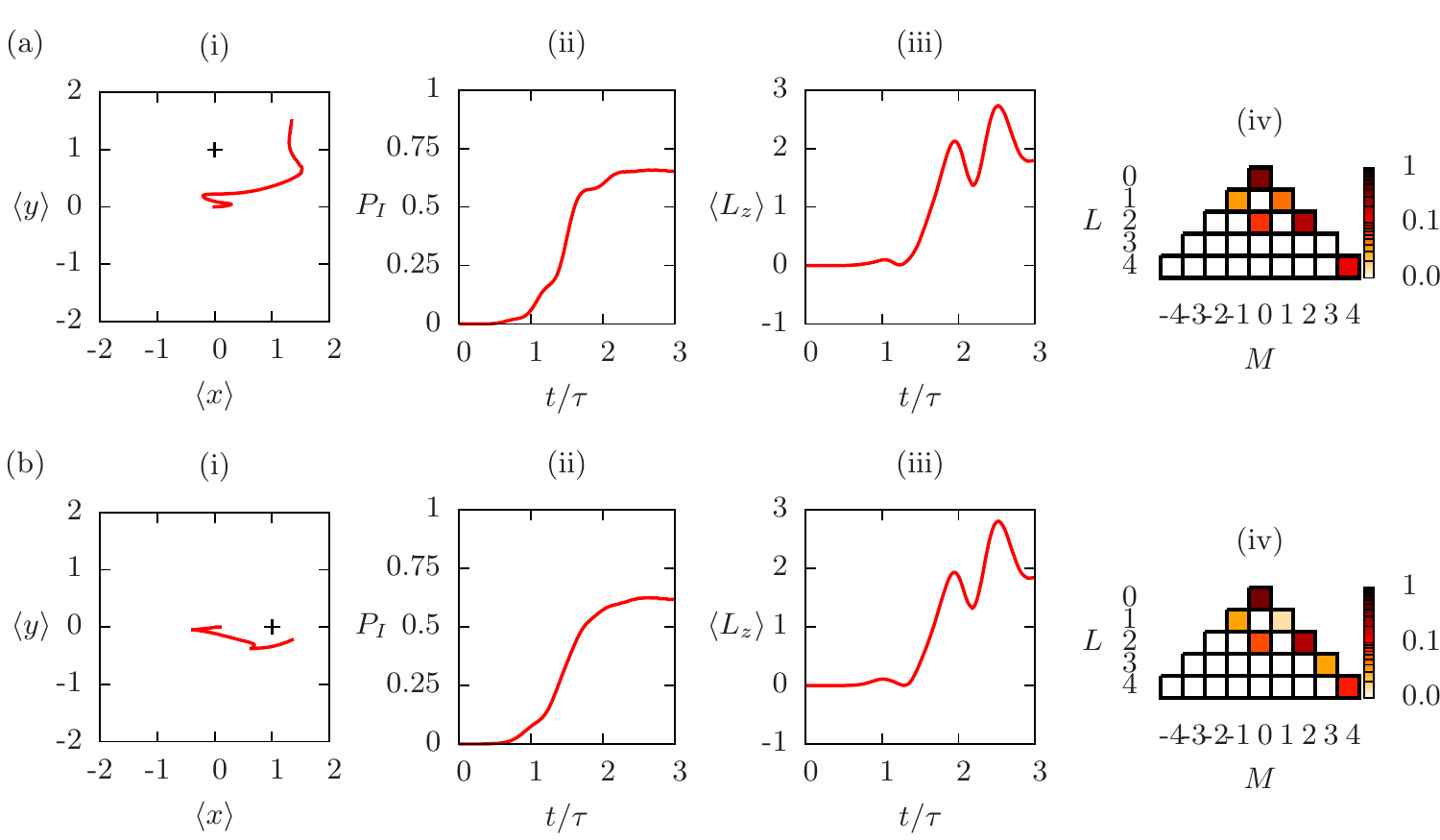}
\end{center}
\caption{\label{fig:NotCentered}
(colour online) Results of a simulation with an LG pulse ($\ell=1$, $p=0$) linearly polarized in the $x$ direction with electric field amplitude $A_{0} \, \omega = 1.4 \cdot 10^{4}$~au, for the light vortex placed at (a) $(x,y)=(0,1)$~au and (b) $(x,y)=(1,0)$~au.
(i) Expected value of the electron in the $xy$ plane, $(\langle x \rangle, \langle y \rangle)$. The cross indicates the position of the vortex.
Temporal evolution of (ii) the ionization probability $P_{\rm I}$ and (iii) the expected value of the angular momentum along the $z$ axis, $\langle L_z \rangle$.
(iv) Projection of the electron quantum state at the end of the pulse into spherical harmonics $Y_L^M$. }
\end{figure}

In this section we consider again a linearly polarized $\ell=1$ LG pulse, as in \sref{sec:LGLinear}, but taking into account a small displacement of the light vortex, while keeping the hydrogen position at the origin.
The results of a simulation where the vortex is located at $(x,y)=(0,1)$~au are plotted in \fref{fig:NotCentered}(a).
From the projected density of the electron wavefunction is difficult to perceive the transverse displacement of the electron, so we plot the mean value of $x$ and $y$ in \fref{fig:NotCentered}(a-i) to study the motion of the electron.
The expected value of the electron position shows that the motion begins oscillating in the polarization axis until it is far enough from the nucleus and then starts rotating counterclockwise around the vortex.
In this case, at the end of the pulse, $P_{\rm I} = 62$ \% and $\langle L_z\rangle = 1.86$~au, see figures~\ref{fig:NotCentered}(a-ii) and \ref{fig:NotCentered}(a-iii), both greater than in the case where the vortex is located at the origin, because now the atom starts at a position where the electric field is more intense.

On the other hand, \fref{fig:NotCentered}(b) shows the results for the light vortex located at $(x,y)=(1,0)$~au. 
We can see a clear mean motion of the electron too, mostly in the polarization direction but the OAM breaks again the symmetry making the electron rotate counterclockwise around the vortex.
As in the previous case, at the end of the pulse we obtain larger values for $P_{\rm I}$ (65 \%) and $\langle L_z\rangle$ (1.79~au) than in the case where the vortex is located at the origin.
The analysis of the spherical harmonic spectrum of the excited electron state is shown in figures~\ref{fig:NotCentered}(a-iv) and \ref{fig:NotCentered}(b-iv).
Comparing these projections with the ones in the case of the optical vortex placed at the origin, see \fref{fig:LGLinear}(d), we see the appearance of a small contribution of $Y_{1}^{1}$, $Y_{1}^{-1}$ and $Y_{3}^{3}$, showing the excitation of electronic transitions with $\Delta L = 1$ that, at first glance, seem to be in disagreement with the selection rules.
The selection rules from \sref{sec:SelectionRules} are calculated for the atom located at the center of the optical vortex. Thus, to address the problem appropriately, we would need to rewrite the vector potential in terms of LG modes centered at the center of the atom in order to take into account the displacement of the light vortex.
This would lead to components with $\ell = 0$, and thus, to the appearance of the electronic transitions with $\Delta L = 1$.
Note that an analogous interpretation could be made by writing the electron state spherical harmonics centered at the vortex position.

The study of the full dependence of the absorption of OAM on the relative position between atom and light vortex is computationally very demanding and it is out of the scope of this paper.
However, the small displacement of 1~au here considered has been enough to show that electronic excitations with large angular momentum take place even if the electron is not on the optical vortex. 
In any case, increasing the distance between the atom and the light vortex up to a point that the azimuthal dependence of the field becomes negligible, it is reasonable to expect that the standard selection rules for plane waves would be recovered.

\subsection{Longer pulses}\label{sec:LongPulses}

\begin{figure}
\begin{center}
\includegraphics{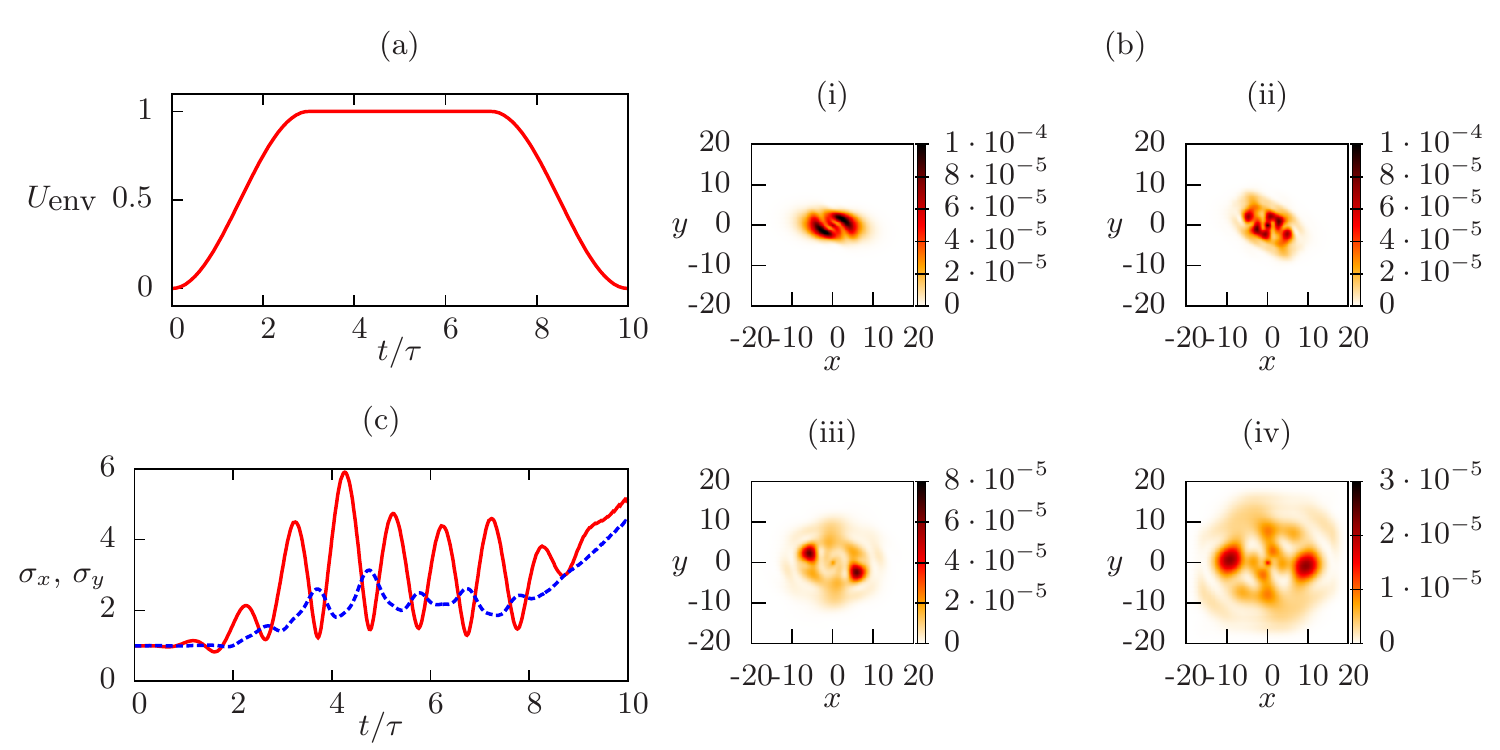}
\end{center}
\caption{\label{fig:LongPulse}
(colour online) Results of a simulation with an LG pulse ($\ell=1$, $p=0$) 10 cycles long linearly polarized in the $x$ direction with electric field amplitude $A_{0} \, \omega = 1.4 \cdot 10^{4}$~au.
(a) Temporal envelope of the pulse.
(b) Projection of the excited state $\vert \delta \psi (t)\rangle$ onto the plane $xy$ at times (i) $t = 3 \tau$, (ii) $t = 8 \tau$, (iii) $t = 9 \tau$ and (iv) $t = 10 \tau$
(c) Temporal evolution of the standard deviations in the $x$ (solid red curve) and $y$ (dashed green curve) directions of the electron distribution $\vert\psi(\vec{r},t)\vert^2$.}
\end{figure}

In this section, we consider the same linearly polarized $\ell = 1$ LG pulse as in \sref{sec:LGLinear}, but with a pulse 10 cycles long.
Thus, the sine squared envelope in~\eref{Eq:VectorPotential} is substituted by:
\begin{eqnarray} \label{Eq:LongEnvelope}
\eqalign{
U_{\rm env}(t) =
\cases{
\sin^2 \left( \pi t / (6 \tau) \right) & \quad \textrm{for $t < 3 \tau$} \\
1 & \quad \textrm{for $3 \tau \leq t \leq 7 \tau$ }  \\
\sin^2 \left(\pi (10 \tau - t) / (6 \tau) \right) & \quad \textrm{for $t > 7 \tau$}
} , }
\end{eqnarray}
i.e., the temporal envelope starts with a smooth sine squared function, which reaches its maximum at three cycles, and afterwards the temporal envelope is constant at its maximum value, until the last three cycles when it ramps down, see \fref{fig:LongPulse}(a).

In \fref{fig:LongPulse}(b), we plot the projection at different times during the pulse: after 3 (when $P_{\rm I} = 50 \%$), 8 ($P_{\rm I} = 67 \%$), 9 ($P_{\rm I} = 66 \%$) and 10 cycles ($P_{\rm I} = 65 \%$).
As we can see, after three cycles, even though the electron strongly interacts with the light field, the excited state remains trapped in the transverse plane, due to its ponderomotive force.
In \fref{fig:LongPulse}(c) we plot the standard deviations $\sigma_x = \langle x^2 - \langle x \rangle^2 \rangle$ and $\sigma_y = \langle y^2 - \langle y \rangle^2 \rangle$.
$\sigma_x$ and $\sigma_y$ oscillate until the pulse intensity decreases, and then the electron cloud starts expanding linearly on both the $x$ and $y$ axis. This can also be seen on the last three snapshots of \fref{fig:LongPulse}(b).

At the end of the simulation, 30\% of the wavefunction is lost in the absorbing boundaries, due to the long interaction.
Although the wavefunction cannot expand in the transverse plane, it expands along the propagation direction and reaches the absorbing grid boundary at $z= \pm 12.5$~au.
Note that in the previous simulations the decrease of the norm at the end of the pulse was at most 2 \%.
	
\section{Conclusions}\label{sec:Conclusions}

We have investigated in detail the transfer of angular momentum to an atom interacting with an ultrashort and ultraintense light pulse carrying both orbital angular momentum and spin. In particular, we have performed 3D numerical simulations of a hydrogen atom, initially in its electron ground state, coupled to an LG pulse for different polarization states, carrier-envelope phases, and relative positions between the atom and the light vortex. We have focused on the photoionization problem by choosing the light carrier frequency to be $\omega = 1$~au well above the ionization threshold of 0.5~au.

The temporal evolution of the electron wavefunction has been projected into spherical harmonics obtaining perfect agreement with selection rules for light-matter interactions including the OAM of light. Notably, we have shown that it is possible to exchange more than one unit of $\hbar$ of angular momentum along the propagation direction of the pulse fulfilling $\Delta{M} = \ell + s$ where $\ell$ is the topological charge of the pulse and $s= \pm 1$ its circular polarization contribution.
For LG pulses, the `local' carrier-envelope phase is position dependent due to the LG azimuthal phase.
Accordingly, we have checked that the photoionization probability and the transfer of angular momentum does not change significantly for different values of the `global' carrier-envelope phase for linearly polarized LG pulses.
We have also investigated the electron dynamics as a function of the relative position between the vortex and the hydrogen atom. For small relative distances, e.g., a few au, the main dynamics are quite similar although small changes in the electronic transition excitations have been observed. Hence, we expect that far from the vortex position, the atom will not absorb OAM since it will interact with a plane-shape-wave field.
Although the light intensity that this theoretical study requires is very large, its value near the pulse vortex position is very small.
From the practical point of view, the generation of light vortices with short pulses and high intensities requires the use of phase holograms and achromatic set ups, as it has been reported recently in experiment~\cite{SolaHigh2008}.

To obtain physical insight into the dynamics of the electron wavefunction and the absorption of angular momentum, we have also computed the de Broglie--Bohm quantum trajectories, plotting for each scenario the dynamics of the electron in the polarization plane.
The mean value of the position given by the trajectories coincides with the one obtained in the orthodox formulation of quantum mechanics, but the collection of trajectories gives a very different view of the process.
See, for instance, the case of the combination of OAM and circular polarization described in \fref{fig:LGRCircular}, where the trajectories clearly rotate around the light vortex, while the mean position remains fixed.

Finally, we would like to remark that the work presented here extends beyond the photoionization scenario and could be used to investigate transitions forbidden by standard selection rules in the electric and magnetic dipole approximation.
Moreover, the generation of OAM laser beams with high intensity is required to excite transitions with large angular momentum exchange, opening the door to new applications ranging from high-harmonic generation with OAM~\cite{JoseA2009,Padgett1997} in attosecond science to nuclear spectroscopy~\cite{Keitel,Leite}.

\ack
The authors acknowledge R. Corbal\'an and X. Oriols for fruitful discussions.
We acknowledge support by the Spanish Ministry of Education and Science under contracts FIS2008-02425, FIS2009-09522, FIS2007-29091-E, and Consolider projects SAUUL and QOIT, CSD2007-00013, CSD2006-00019, and the Catalan, Junta de Castilla y Le{\'o}n, and Junta de Castilla-La Mancha Governments under contracts SGR2009-00347, SA146A08, and PCI08-0093. A. B. acknowledges financial support through grant AP 200801275 from MICINN (Spain).

\appendix

\section{Selection rules derivation}

In this appendix we derive the selection rules corresponding to the interaction Hamiltonian $\mathcal{H}_I$.
The derivation we present here does not differ much with the one in~\cite{OptExPicon2010}, but it includes the light circular polarization which then appears explicitly in the selection rules.

We start by writing the transition probability amplitude as 
\begin{eqnarray}
\langle \psi_\rmf \vert \mathcal{H}_I \vert \psi_\rmi \rangle = -\frac{q}{2m} \langle \psi_\rmf \vert (\bfp\cdot\bfA+\bfA\cdot\bfp) \vert \psi_\rmi \rangle =-\frac{e}{\rmi\hbar}\Delta E\langle \psi_\rmf \vert \bfr\cdot\bfA \vert \psi_\rmi \rangle . \label{eq:CouplingrA}
\end{eqnarray}
by using that $\bfp= - \rmi m [\bfr,\mathcal{H}_0]/\hbar$, where $\Delta E$ is the energy difference between states $\vert \psi_\rmi \rangle$ and $\vert \psi_\rmf \rangle$.
For an $s$-circularly polarized light beam carrying $\ell \hbar$ units of OAM, see \eref{Eq:VectorPotential}, and assuming the dipolar approximation ($\lambda \gg a_{0}$) and the transverse spatial approximation ($w_{0} \gg a_{0}$), the atom placed in the light vortex and ignoring the temporal envelope, one can write (in spherical coordinates)
\begin{eqnarray}
\bfr\cdot\bfA =A_{0}\sqrt{\frac{2 \, (\vert\ell\vert+p)!}{\pi \, p!}}\frac{1}{\vert\ell\vert!}\left(\frac{\sqrt{2}}{w_{0}}\right)^{\vert\ell\vert} r^{\vert\ell\vert+1}\sin^{\vert\ell\vert+1}\theta \, \rme^{\rmi(\ell+s)\phi}\rme^{-\rmi\omega t}+c.c. \label{eq:rA}
\end{eqnarray}
Thus, the angular part of the transition probability amplitude between states $\psi_{\rm i}(\vec{r})=U_{\rm i}(r) \, Y_{\Li}^{\Mi}(\theta,\phi)$ and $\psi_{\rm f}(\vec{r})=U_{\rm f}(r) \, Y_{\Lf}^{\Mf}(\theta,\phi)$ (being $Y_L^M(\theta,\phi)$ the spherical harmonics with the quantization axis in the light propagation direction) obtained by inserting~\eref{eq:rA} into~\eref{eq:CouplingrA} gives two contributions, proportional to the integrals:
\begin{eqnarray}
I_{1}(\ell, s) = & \int Y_{\Lf}^{*\Mf}(\theta,\phi) \, \sin^{\vert\ell\vert+1}\theta \, \rme^{\rmi(\ell+s)\phi} \, Y_{\Li}^{\Mi}(\theta,\phi)d\Omega , \label{Integral1} \\
I_{2}(\ell, s) = & \int Y_{\Lf}^{*\Mf}(\theta,\phi) \, \sin^{\vert\ell\vert+1}\theta \, \rme^{-\rmi(\ell+s)\phi} \, Y_{\Li}^{\Mi}(\theta,\phi)d\Omega . \label{Integral2} 
\end{eqnarray}
\Eref{Integral1} is the term related with $\rme^{-\rmi \omega t}$ in~\eref{eq:rA} and is associated with the absorption of photons~\cite{Faisal}.
Analogously, \eref{Integral2} is related with $\rme^{+\rmi \omega t}$ and is associated to the stimulated emission of photons.
It is straightforward to see that  $I_{1}(\ell, s) = I_{2}(-\ell, -s)$, i.e., the transitions allowed by the absorption of a $(\ell, s)$ photon are the same as the transitions allowed by the emission of a $(-\ell, -s)$ photon.

The allowed transitions between spherical harmonics given by integral \eref{Integral1} can be calculated by writing $\sin^{\vert\ell\vert+1}\theta \, \rme^{\rmi(\ell+s)\phi}$ in terms of spherical harmonics and evaluating for which transitions $I_1$ does not vanish (see Complements B$_{\rm X}$ and C$_{\rm X}$ in~\cite{Cohen-Tannoudji}).
Thus, defining $\Delta L \equiv \Lf - \Li$ and $\Delta M \equiv \Mf - \Mi$, the absorption selection rules given by \eref{Integral1} are
\begin{eqnarray}
\vert\Delta L\vert \leq \vert\ell\vert + 1 \leq \Li + \Lf , \quad  \Delta M = \ell + s, \quad  \Delta L + \ell \hspace{0,2cm} \textrm{is odd} .
\end{eqnarray}

Analogously, the emission selection rules calculated from \eref{Integral2} are
\begin{eqnarray}
\vert\Delta L\vert \leq \vert\ell\vert + 1 \leq \Li + \Lf , \quad \Delta M = - \ell - s, \quad  \Delta L + \ell \hspace{0,2cm} \textrm{is odd} .
\end{eqnarray}

\section{de Broglie--Bohm formulation}

In this appendix give a brief overview of the main equations of de Broglie--Bohm formulation of quantum mechanics \cite{Bohm-1,Bohm-2,Bohm-3}.
By using $\bfp = - \rmi \hbar \bfN$, we can rewrite Schr\"odinger equation \eref{Eq:Schrodinger} as:
\begin{eqnarray}
\eqalign{
\rmi \hbar \frac{\partial \psi}{\partial t} = &- \frac{\hbar^2}{2m} \bfN^2 \psi + \frac{\rmi \hbar q}{m} \bfA \cdot \bfN \psi + \frac{\rmi \hbar q}{2m} \left(\bfN \cdot \bfA \right) \psi \\ &+ \frac{q^2}{2m} \bfA^2  \psi + qV \psi  .} \label{Eq:App-Schro}
\end{eqnarray}
Then, by casting into it the polar form of the wavefunction, $\psi = R\rme^{\rmi S / \hbar}$, and separating real and imaginary parts, in a similar manner as done in~\cite{Bohm-1}, we obtain
\begin{eqnarray} 
\frac{\partial S}{\partial t} = & \frac{\hbar^2}{2m} \frac{\bfN^2 R}{R}  -\frac{1}{2m} \left(\bfN S - q \bfA\right)^2 - qV , \label{Eq:App-QHJeq} \\
\frac{\partial R^2}{\partial t} = & - \frac{1}{m} \bfN \left[ R^2 \left( \bfN S - q \bfA \right) \right] . \label{Eq:App-Conteq}
\end{eqnarray}
\Eref{Eq:App-QHJeq} is the so-called quantum Hamilton--Jacobi equation because of its similarity with the (classical) Hamilton--Jacobi equation but with one additional term, the quantum potential, accounting for the quantum features of the system. This similarity suggests the definition of the particle velocity as:
\begin{eqnarray}
\frac{\rmd}{\rmd t} \bfx_k[t] = \bfv_k[t] = \frac{1}{m} \left.\left( \bfN S - q\bfA \right) \right|_{(\bfx_k[t], t)}. \label{Eq:App-BohmVel}  
\end{eqnarray}
Thus, \eref{Eq:App-Conteq} becomes a continuity equation ensuring that the trajectories distribution is given by $R^2(\bfr, t)$ at all times. 

\begin{figure}
\begin{center}
\includegraphics{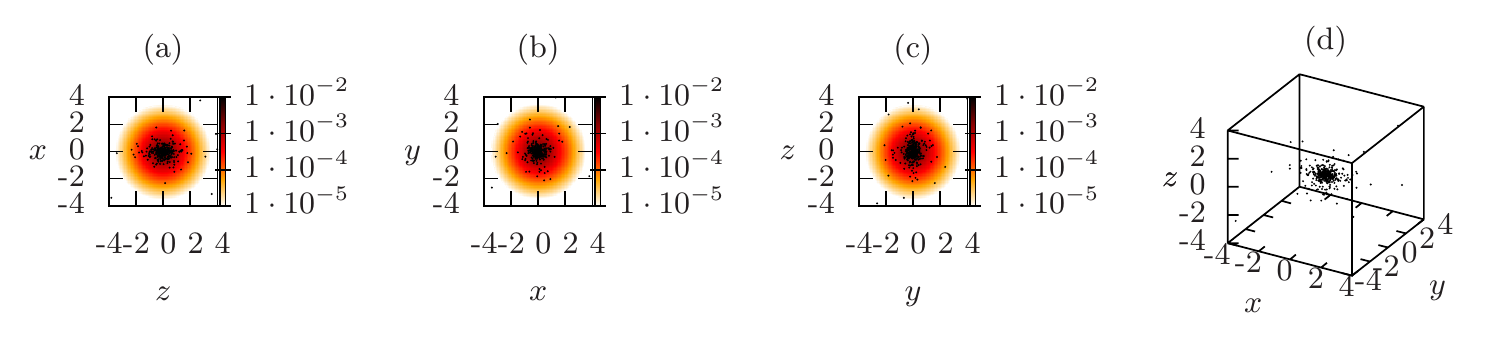} 
\end{center}
\caption{\label{fig:Distr0}
(colour online) Initial positions of the trajectories, on the planes (a) $zx$ (b) $xy$ and (c) $yz$, and (d) in space. \label{fig:app-distr0}}
\end{figure}

The initial positions of trajectories $\left\{ \bfx_k[t_0] \right\}$ are distributed randomly following the probability density function of the hydrogen atom ground state: $R^2(\bfr, t_0) = \frac{1}{\pi a_0^3} \rme^{-2r/a_0}$, see \fref{fig:app-distr0}. Then, after solving \eref{Eq:App-Schro} (or, alternatively, \eref{Eq:App-QHJeq} and \eref{Eq:App-Conteq}) we find the quantum trajectories, $\left\{ \bfx_k[t] \right\}$, (time evolution of the positions) by integrating \eref{Eq:App-BohmVel}:
\begin{eqnarray}
\bfx_k[t] = \int_{t_0}^t \bfv_k[t] \rmd t . \label{Eq:App-XT}
\end{eqnarray}

\end{document}